\documentclass[preprint]{aastex}
\usepackage[english]{babel}
\usepackage{graphicx}
\usepackage{subfigure}
\usepackage{placeins}

\usepackage{array}   %% for the table
\usepackage{booktabs} %% table
\usepackage{multirow} %%table

\usepackage{multicol}
\usepackage{adjustbox} %%to fit the table into a column of the page

\usepackage{natbib}

\newcommand{\aftr}{\textbf}

\shorttitle{Time-resolved emission from bright hot pixels }
\shortauthors{Djorgovski et al.}

\begin{document}

\title{EUV flickering of \aftr{solar} coronal loops: a new diagnostic of coronal heating}

%Space dependence of EUV emission from bright hot pixels of an active region observed with SDO/AIA and multi-stranded loop modeling}

\author{E. Tajfirouze\altaffilmark{1}, F.Reale\altaffilmark{1,2}, G. Peres\altaffilmark{1,2}, P. Testa\altaffilmark{3}}
\affil{\altaffilmark{1}Dipartimento di Fisica e Chimica, Universit\`a di Palermo, Piazza del Parlamento 1, 90134, Italy}
\affil{\altaffilmark{2}INAF-Osservatorio Astronomico di Palermo ``G.S. Vaiana'', Piazza del Parlamento 1, 90134, Italy}
\affil{\altaffilmark{3}Harvard-Smithsonian Center for Astrophysics, 60 Garden Street, Cambridge, MA 02138, USA}
\email{reale@astropa.unipa.it}

\begin{abstract}
A previous work of ours found the best agreement between EUV light curves observed in an active region core (with evidence of super-hot plasma) and those predicted from a model with a random combination of many pulse-heated strands with a power-law energy distribution. We extend that work by including spatially resolved strand modeling and by studying the evolution of emission along the loops in the EUV 94 \AA\ and 335 \AA\ channels of the Atmospheric Imaging Assembly on-board the Solar Dynamics Observatory. Using the best parameters of the previous work as the input of the present one, we find that the amplitude of the random fluctuations driven by the random heat pulses increases from the bottom to the top of the loop in the 94 \AA\ channel and, viceversa, from the top to the bottom in the 335 \AA\ channel. This prediction is confirmed by the observation of a set of aligned neighbouring pixels along a bright arc of an active region core. Maps of pixel fluctuations may therefore provide easy diagnostics of nano-flaring regions.
\end{abstract} 


\section{Introduction}
\label{sec:intro}

The quest for coronal heating mechanisms is a difficult one for several concurring reasons. A very important one is that the corona is finely structured and the elementary fibrils where elementary heating releases may occur cannot be resolved at present times.  Indirect evidence is therefore fundamental. 
Evidence for small amounts of very hot plasma ($> 4$ MK) has been detected in active region cores \citep{McT09,Rea09,Re09,Tes11,Mic12,Te12,Bro14,Pet14,Cas15}. The presence of such hot plasma and of a cooler steady state emission is a possible indication of impulsive heating events on small spatial scales, but further support is necessary. It might be also the case that different heating mechanisms play a role \citep[e.g.][]{Rea14}. It would then be very important to pinpoint where some mechanisms are in action and wherelse others.

Models of multi-stranded pulse-heated loops have been used to explain the evidence of super-hot plasma  and to constrain the heating release \citep{Par88,Car94,Kli08,Via12,car12, Ree13, car14, Anl14,Kob14,Cas15,Lop15,Tam15}

In a previous work \citep[][hereafter Paper I]{Taj15}, the analysis of observations in the EUV band has shown that a storm of heat pulses in a coronal loop is able to explain steady but flickering light curves in single spatial elements. A 0D loop model \citep[EBTEL,][]{Kli08,car12} that computes average loop quantities with no spatial resolution was used with success to produce a grid of simulations of pulse-heated strands with different heating rates and two values for the duration, namely 50 s and 500 s, the latter comparable to plasma cooling times. We assumed a power-law energy distribution of the pulses and three possible numbers of independent strands, i.e. 10, 100 and 1000. In the assumption of active region loops made of bundles of strands, we then combined the pulse-heated model strands with random energy and random start times of the heat pulses into 10000 realisations for each set of parameters.  An artificial intelligence method allowed to find a realisation that best reproduces the observation in two EUV channels. The combination of parameters for this realisation is a power law with index 1.5, the short pulse duration (50 s), and the largest number of strands (1000).
The slope 1.5 can reproduce the observed active region core. Other studies obtain steeper slopes \citep[e.g. 2.5;][]{Lop15}. We remark that the shallow law is related to a narrow energy range ($\sim 1$ decade) and to a very localised and homogeneous region. 

In this work we use the best realisation of Paper I as reference model to investigate a more detailed issue related to the active region coronal loops. We study how the temporal evolution of the emission changes with the spatial location along the loops. To do this, we need a model with spatial resolution; thus, we move from 0D to 1D loop modeling. 

With respect to the previous work, we address a new question: can the multi-strand pulse-heated loop model match also the spatial trends of EUV emission variability? Therefore, the aim of this work is to use the 1D model -- building on the 0D model result --to find a rather new answer, unreachable with 0D models. The answer will lead us to develop a new diagnostic tool, based on the trends of the EUV emission fluctuations.
In the light of this, we freeze the model parameters to those of the previous work and explore the spatial dependence determined by this parameter combination, and possible diagnostics that we will compare to the observation.

\section{The analysis and results}

We use a one-dimensional time-dependent loop model \citep{Per82,Bet01} which considers the magnetic field as static and solves the hydrodynamic equations (for a toroidal geometry with constant cross section) along one representative strand, using an assumed heating rate. The model includes the effects of gravity for a curved geometry,  thermal conduction and radiative losses for an optically thin plasma, and compressional viscosity.
The differential fluid equations of conservation of mass, momentum and energy are approximated with finite difference equations and are solved on an adaptive grid, the number and the spatial size of which is modified at each step of time integration to resolve properly the steep transition region, especially at the beginning of a flare. The code allows for including  both the spatial and temporal dependencies in the heating term. 
As in Paper I, we let a single strand evolve under the effect of a heat deposition uniformly distributed along the strand. The details of the heat deposition are important in the initial phases of the evolution, but much less over the longer time scales of our interest, when the plasma quickly loses memory of the previous evolution. The model computes the evolution of the density, temperature, and velocity of the plasma along the single coordinate of the loop.

The initial cool and tenuous atmosphere is kept steady by a low constant heating of  $0.23 \times{10}^{-4}\ erg\ cm^{-3}s^{-1}$  which supports a pressure of $0.018\ dyn\ cm^{-2}$ and a temperature of $0.62\ MK$ in a  semicircular loop of half length $\sim2.5\times{10}^9 \ cm$. These are not exactly the same as those used in Paper I, but the evolution does not change significantly if the initial strand is much cooler and tenuous than after heating, which is our case.

\begin{figure}[!ht]               %%%%%%Figura3%%%%%%%%%%%%%
\centering
 \subfigure[]{\includegraphics[width=8cm]{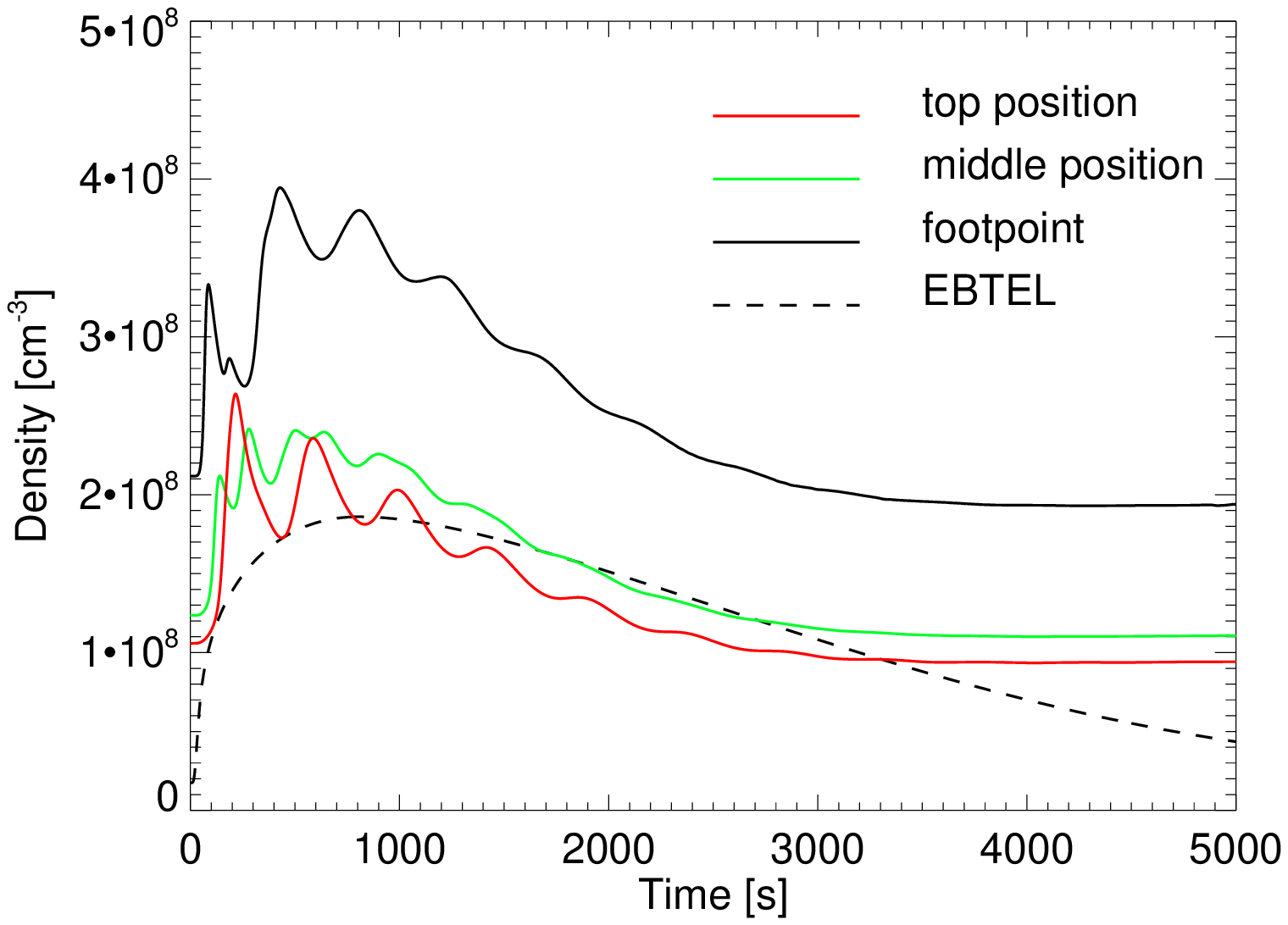}}
 \subfigure[]{\includegraphics[width=8cm]{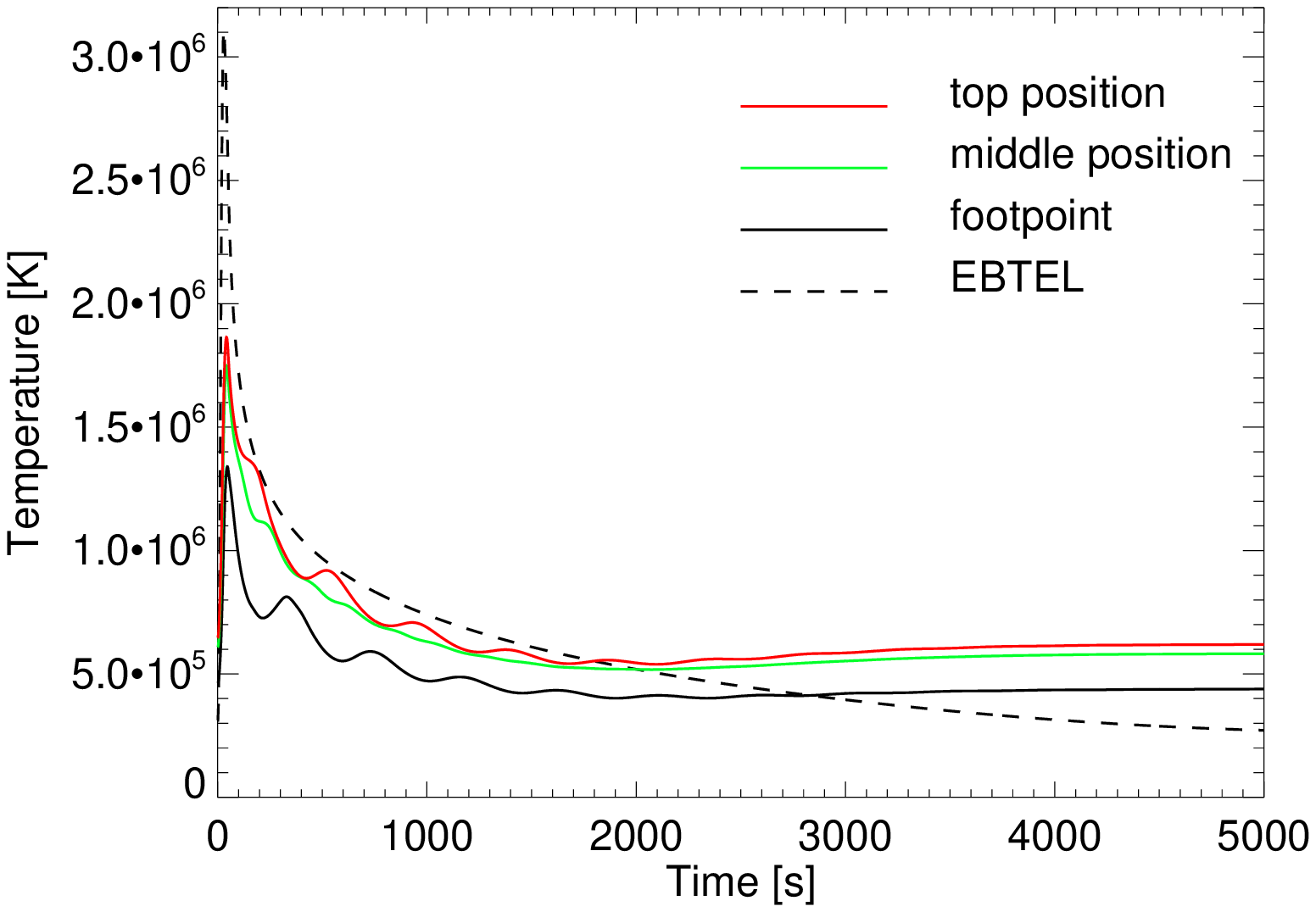}}
 \subfigure[]{\includegraphics[width=8cm]{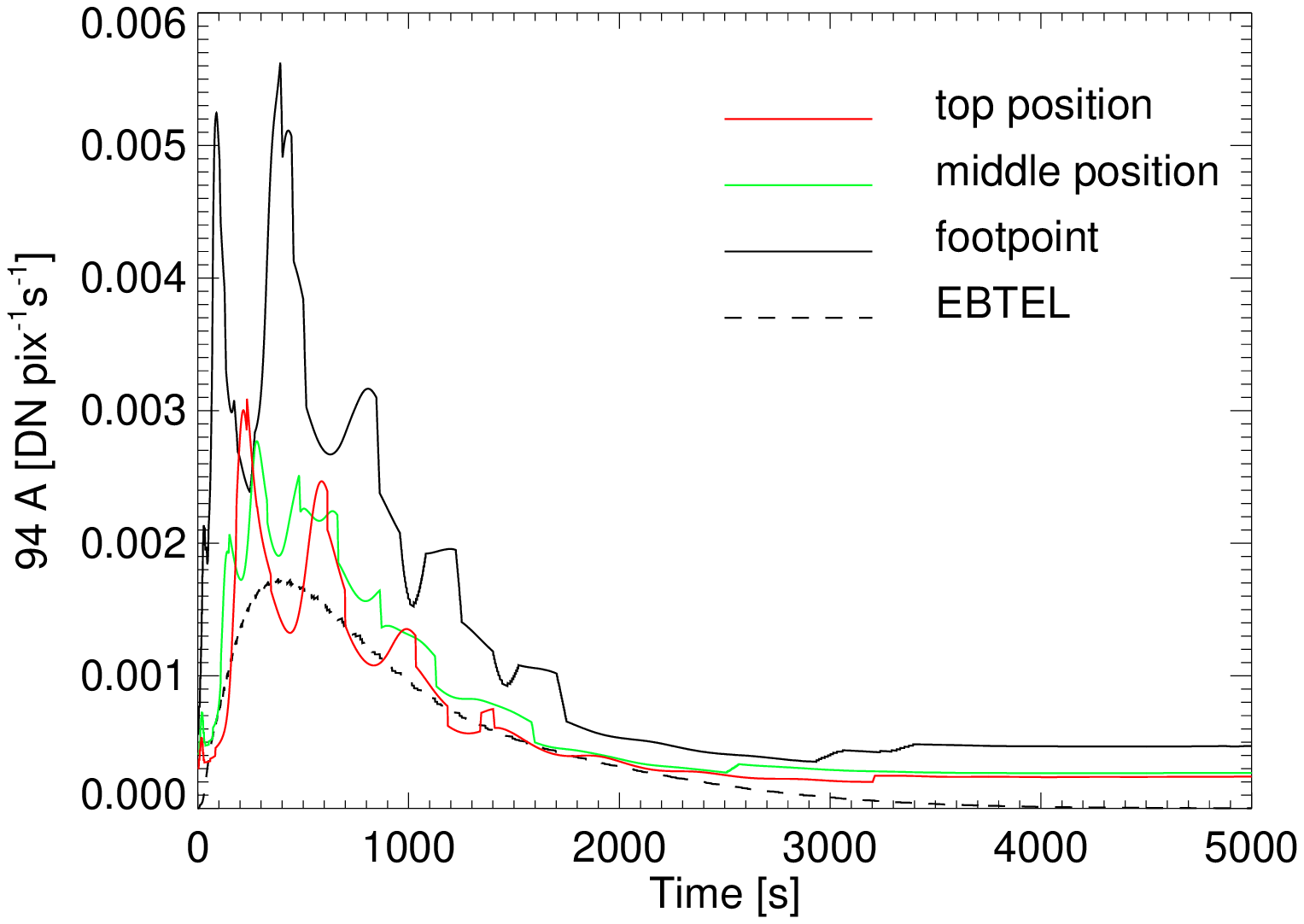}}
 
 \caption{Evolution of (a) Density (b) Temperature (c) and Intensity in 94 ~\AA\ channel  at the three labelled locations along a pulse-heated strand ({\it solid lines}). The evolution obtained with the 0D model is also shown for comparison ({\it dashed lines}).}
 
\label{fig3}
\end{figure}

The evolution of the loop plasma under the effect of a single heat pulse is well known from previous work \citep[e.g.][]{Car94,Car04,Bra06,Rea08,Gua10}.
Figure~\ref{fig3} shows the evolution of the density and temperature at three representative positions (top, middle, footpoint) of a single strand as the relaxed loop atmosphere is perturbed by a heat pulse with a triangular time evolution and a peak intensity of $h=0.003$ erg cm$^{-3}$ s$^{-1}$ and with a total duration of $\tau=50$ s. As expected, the overall evolution is similar to that of the single strand derived in Paper I with the 0D model, which is also shown in Figure~\ref{fig3} for comparison. Again we see a steep temperature rise and a slower decay, and a slower evolution of the density. The evolution is very similar at all positions, except for a shift to higher temperature and to lower density toward the strand top, as expected from temperature and density stratification. Figure~\ref{fig3} also shows periodic fluctuations in both quantities and at all positions that are not present in the average evolution obtained with the 0D model. These are due to pressure waves that travel back and forth along the strand triggered by the highly discontinuous heating. Traveling perturbations cannot be described by the 0D model. The maximum temperature at the top is close to 2~MK, and the density reaches around $5 \times 10^8$ cm$^{-3}$ close to the footpoints. The maximum temperature obtained with the 0D model is higher for this very short heat pulse. This makes a small difference in the overall evolution because the plasma remains so hot for a very short time and this does not influence the subsequent evolution. As in Paper I, we assume that each strand is heated only once by a single pulse whose energy and power are selected at random from a power-law distribution.

Using the outputs of 1D hydrodynamic model we can derive the emission at any time and at single positions along the strand as $n(s,t)^2 G[T(s,t)]$ where $s,t$ are the spatial and temporal coordinates, respectively, $n$ is the density, $T$ is the temperature, $G(T)$ is the instrument sensitivity function to plasma emitting at temperature $T$. As in Paper I, in this work we consider the emission as observed in the 94~\AA\ and 335~\AA\ EUV channels of the Atmospheric Imaging Assembly (AIA) on-board the Solar Dynamics Observatory (SDO). These channels are best sensitive to plasma at $\sim 4-6$ MK, and $\sim 3$ MK respectively, in the core of an active region. Figure~\ref{fig3} also shows the expected emission in the 94 \AA\ channel, which peaks at an intermediate time between the temperature and the density. Although the overall evolution is similar to that obtained with the 0D model (also shown in the figure), also here we see periodic fluctuations that are larger than those of temperature and density, but these will be mostly washed out when we will randomly combine the light curves of a bundle of strands. 

We generate the same grid of strand models for the short-duration heat pulses as we did in Paper I, i.e. the same sampling of the energy range, and we combine them with the same population, distribution and starting times as the most successful realisation in Paper I. As mentioned in Section~\ref{sec:intro}, the energy distribution is a power law with index $\alpha = 1.5$, and we extract 1000 strand models, which we overlap with exactly the same sequence as in Paper I in the time range of 10000~s. Whereas in Paper I we summed the average loop emission of single strands to obtain a single light curve, here we sum the space-resolved emission along the strand, i.e., assuming that all strands are parallel and start and end at exactly the same locations on the solar surface. The plasma distribution is rather smooth along the strands and therefore no significant changes are expected for small spatial shifts of one strand to the other.

\begin{figure}[!ht]               %%%%%%Figura3%%%%%%%%%%%%%
\centering
 \subfigure[]{\includegraphics[width=12cm]{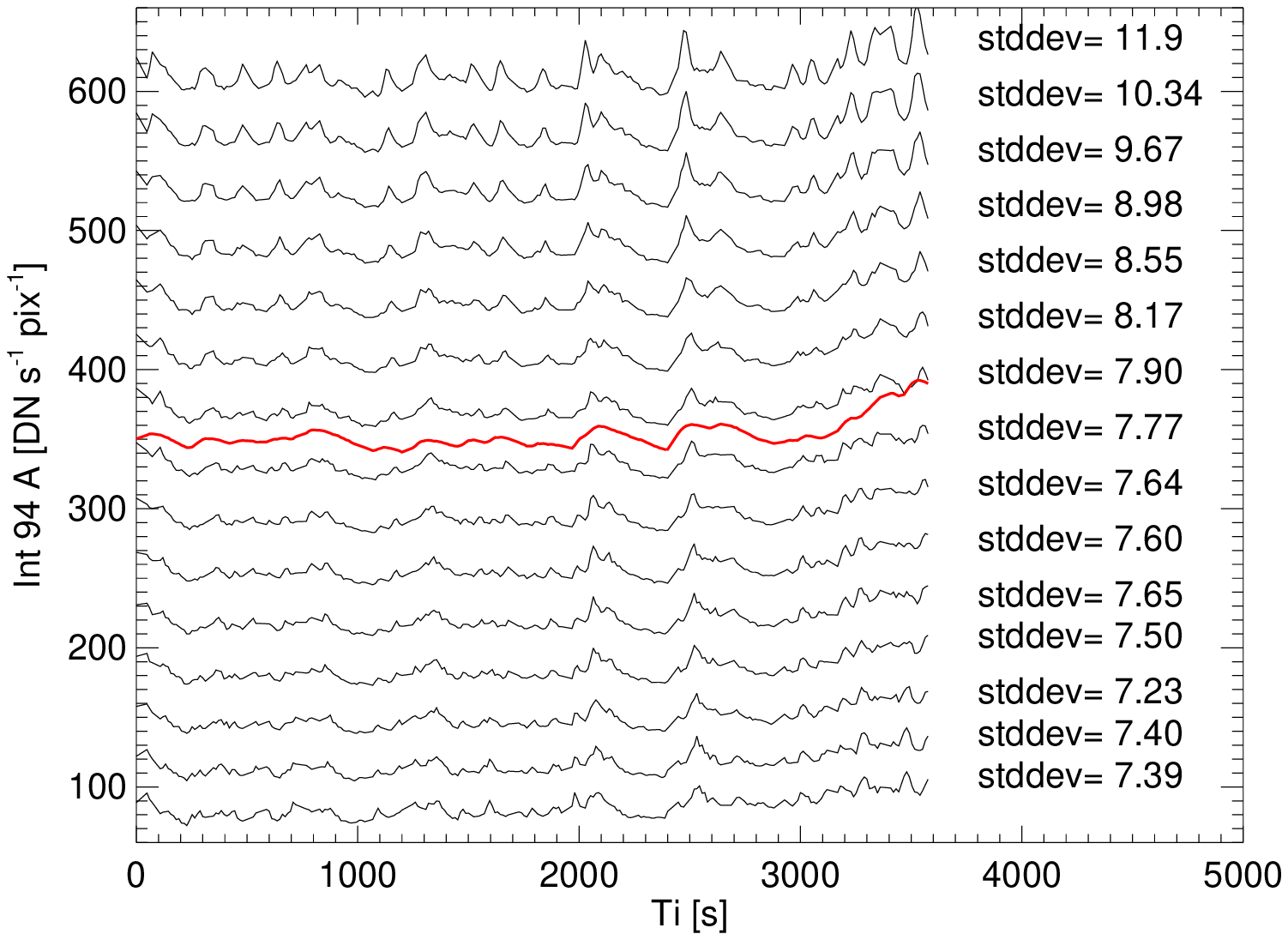}}
 \subfigure[]{\includegraphics[width=12cm]{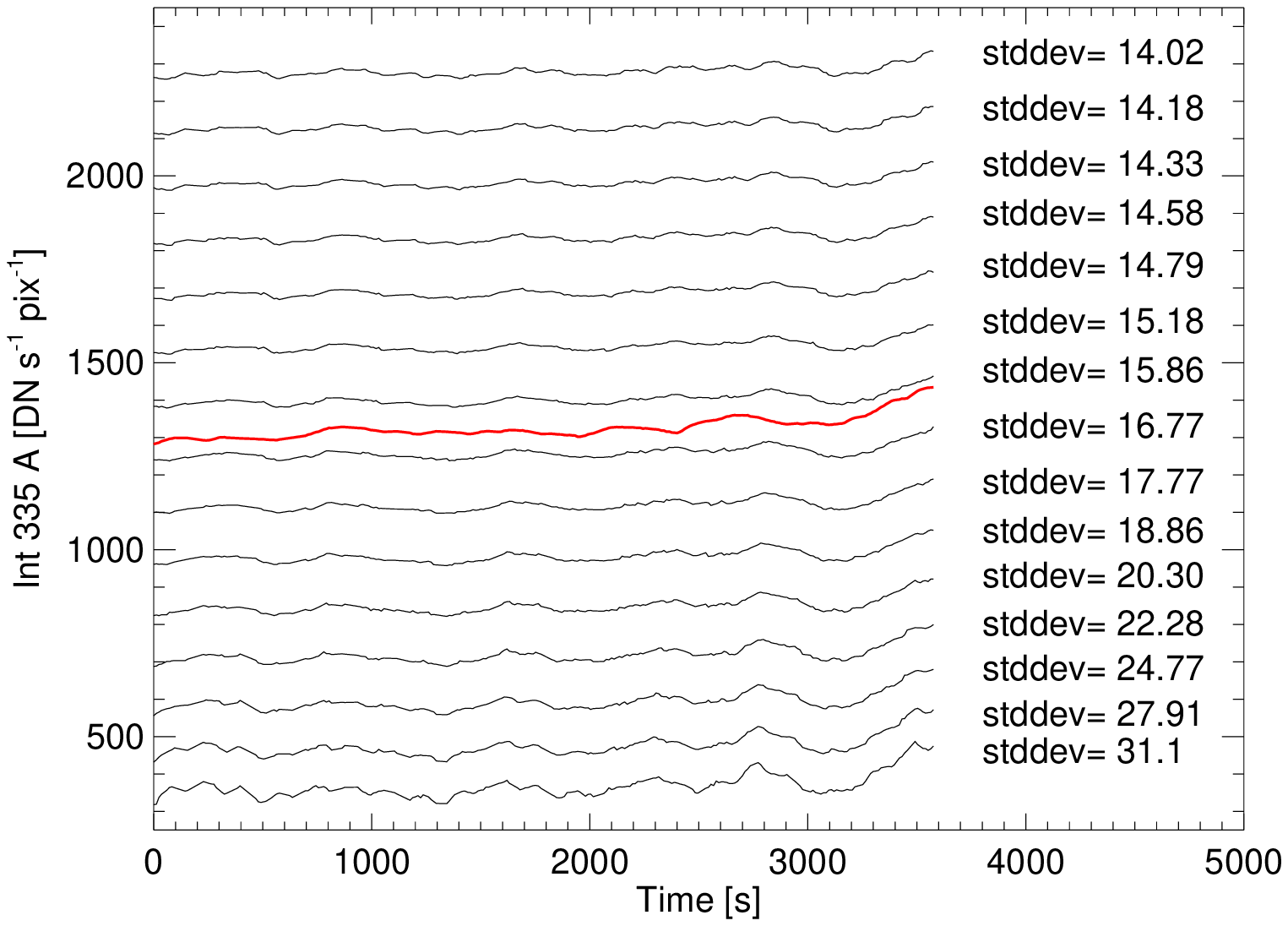}}
 
 \caption{Model light curves along the loop for the 94~\AA\ (left) and 335~\AA\ (right) channel ({\it black lines}). The light curves from the bottom to the top of the plot are for locations from closer to the footpoints to closer to the loop top, and are shifted by a constant positive value from the bottom one (10 and 20 DN s$^{-1}$ pixel$^{-1}$ for the 94~\AA\ and 335 \AA\ channel, respectively). Corresponding values of the standard deviation from the mean are listed on the right of each curve. The light curves obtained with the 0D model are also shown for comparison ({\it red lines}).} 
 
\label{fig4}
\end{figure}

We have derived the total light curves from 1000 heated strands at different positions along the loop. Figure~\ref{fig4} shows the light curves in the 94 \AA\ and 335 \AA\ channels at 15 positions along the loop. For a direct comparison with the observations, we assume an average cross-section factor between those obtained in Paper I (i.e., a strand cross-section of 0.54 pixels). The light curves obtained with the 0D model (see Paper I) are also shown for comparison. As in Paper I the light curves in the 94 \AA\ channels are more variable and show frequent peaks, because this channel is sensitive to hotter plasma, and therefore to the fast heat pulses. Many minor peaks are due to fluctuations like those shown in Fig.~\ref{fig3}c. In the 335 \AA\ channels we see smoother light curves. However, space resolution allows us to notice that the light curves in the same channel are similar from one location to the other along the loop, and show no obvious correlation to the light curves in the other channel. In the 94 \AA\ channel we also clearly see that the amplitude of the fluctuations of the light curves increase moving from the base to the top of the loop. The opposite occurs in the 335 \AA\ channel: the fluctuations increase from the top to the footpoints. This is not only a visual impression; we measured the standard deviation from the average of each light curve and it grows upwards in the 94 \AA\ channel and downwards in the 335 \AA\ channel, thus confirming the visual impression. This means that we expect more flickering of the emission going to the top of the core loops in the 94 \AA\ channel, and to base in the 335 \AA\ channel. We have investigated the reason for this behaviour, which is not obvious because we are looking at the sum of a multitude of contributions. 

\begin{figure}[!t]
\centering
\subfigure[]{\includegraphics[width=8cm]{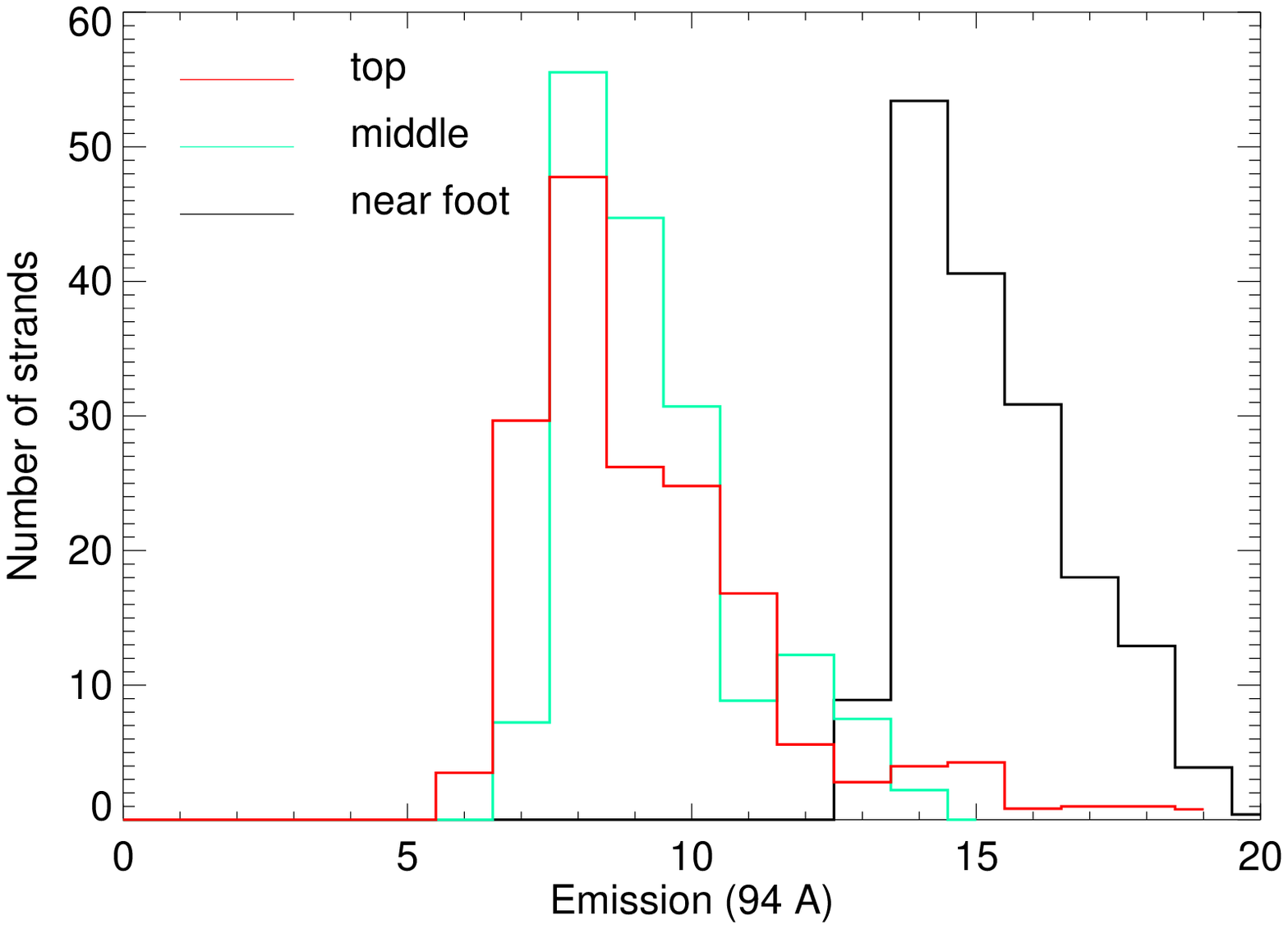}}
\subfigure[]{\includegraphics[width=8cm]{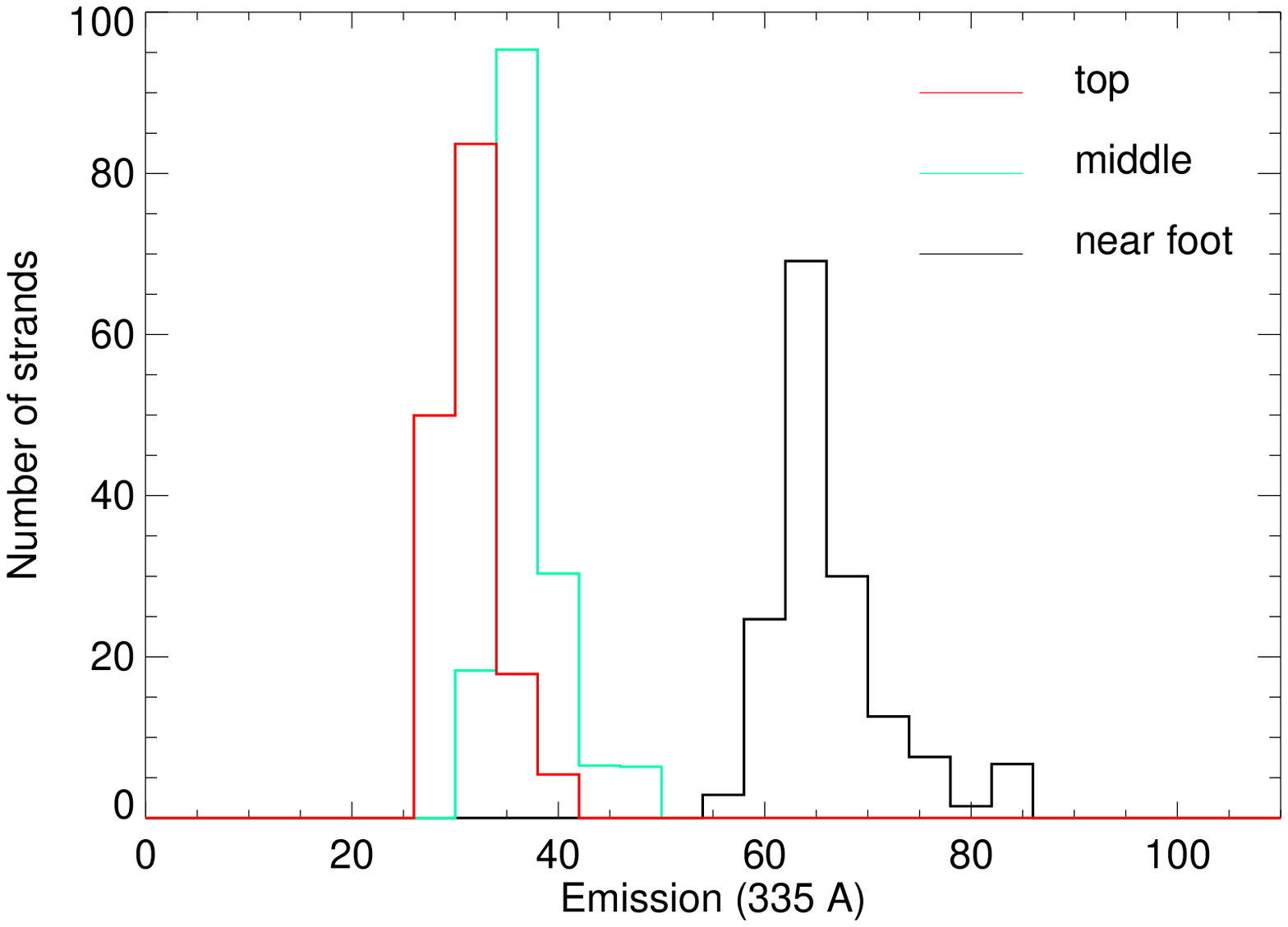}}
\caption{Histograms of the emitting strands vs emission intensity (DN s$^{-1}$ pix$^{-1}$) in the 94 \AA\ channel (a) and 335 \AA\ channel (b) at three different locations along the modelled loop, top (red), middle (green) and footpoints (black). High emission tails are present at the top in the 94 \AA\ channel and at the footpoint in the 335 \AA\ channel.  } 
\label{fig5} 
\end{figure}

A hint can be found examining the brightest strands that contribute to the emission in the two channels. We have checked that the 94 \AA\ channel is sensitive to the hottest plasma at the top of the strands. The hottest strands are also those with the largest emission measure, they are the fewest according to the power law, and they stay very hot for a small time. For these reasons the channel is extremely sensitive to their variations, and in particular to plasma entering and exiting its temperature response at the top of the hottest strands. Since the temperature in each strand decreases downwards to the loop footpoints, the same strands lead to the same effect in the 335 \AA\ channel at the loop footpoints. Fig.~\ref{fig5} shows the histograms of the emitting strand along the loop and confirms that the 94 \AA\ one  has a long tail to the high emission side at the top and not elsewhere, the 335 \AA\ one at the footpoint. These strands in the tail determine the larger fluctuations in the two channels. Since the hottest and brighest strands are those heated by the most intense energy pulses we conclude that the few most intense heating events  ultimately determine this trend of fluctuations.

As a feedback to the data, we have searched for a similar behavior in the observations. We consider the same observation of AR11117 on 2010 October 28 as in \cite{Rea11}, and, in particular, the same sequence of 246 and 228 co-aligned images in the  94 \AA\ and 335 \AA\ channel, respectively, as in Paper I, for a total duration of $\sim 1$ hour. The images are at full space resolution with an exposure time of 2.9 s in both channels. The cadence of the images is 12 seconds, with some gaps up to $\sim$ 1 minute.

%
%\begin{figure*}[!t]
%\centering
%\setlength\fboxsep{0pt}
%\setlength\fboxrule{2pt}
%%\includegraphics[width=12cm]{std_emissionmeasur.ps}
%\includegraphics[width=12cm]{fig_strands.eps}
%%\plotone{fig1.ps}
%%\includegraphics[width=\textwidth,height=4cm]{fig1.ps}
%\caption{Number of strands that contribute to the emission along the loop. We show the brightest 50\% ones in the 94~\AA\ channel (stars) and the lowest 5 \% ones (-440, diamonds) in the 335~\AA\ channel.} 
%\label{fig5} 
%\end{figure*}

%\section{Results and comparison with observations}
%\label{sec:obs}

\begin{figure}[!ht]               %%%%%%Figura3%%%%%%%%%%%%%
\centering
 \subfigure[]{\includegraphics[width=12cm]{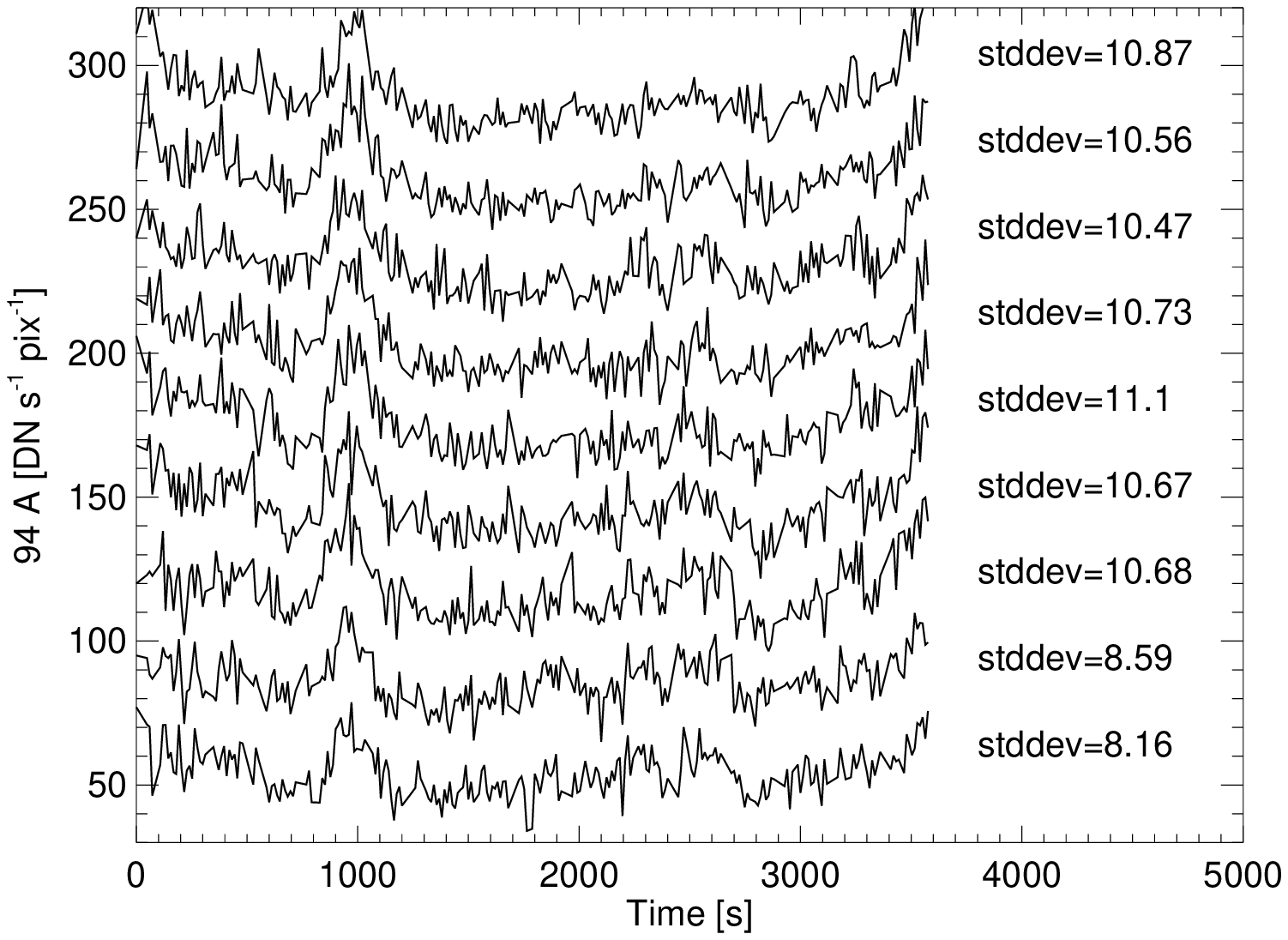}}
 \subfigure[]{\includegraphics[width=12cm]{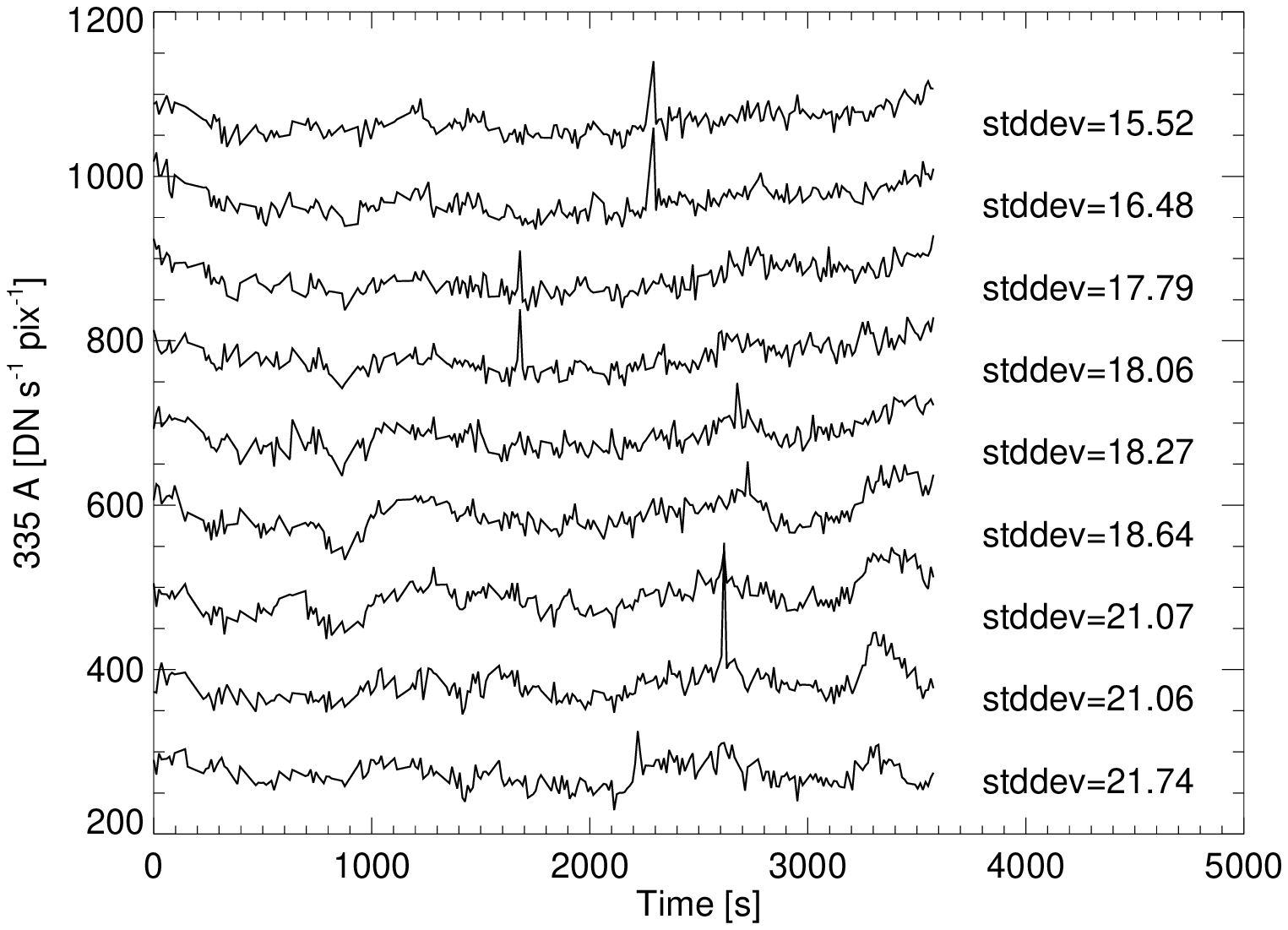}}
 
 \caption{Observed light curves in neighbouring pixels in a row for 94~\AA\ (left) and 335~\AA\ (right) channel. Bottom-up roughly proceeds from closer to the loop footpoints to closer to the loop top, i.e. left to right in Fig.~\ref{fig6}. The light curves are shifted by a constant positive value from the bottom one (30 and 100 DN s$^{-1}$ for the 94~\AA\ and 335 \AA\ channel, respectively). Corresponding values of the standard deviations are reported on the right of each curve. }
 
\label{fig2}
\end{figure}

\begin{figure}[!ht]               %%%%%%Figura3%%%%%%%%%%%%%
\centering
\subfigure[]{\includegraphics[width=8cm]{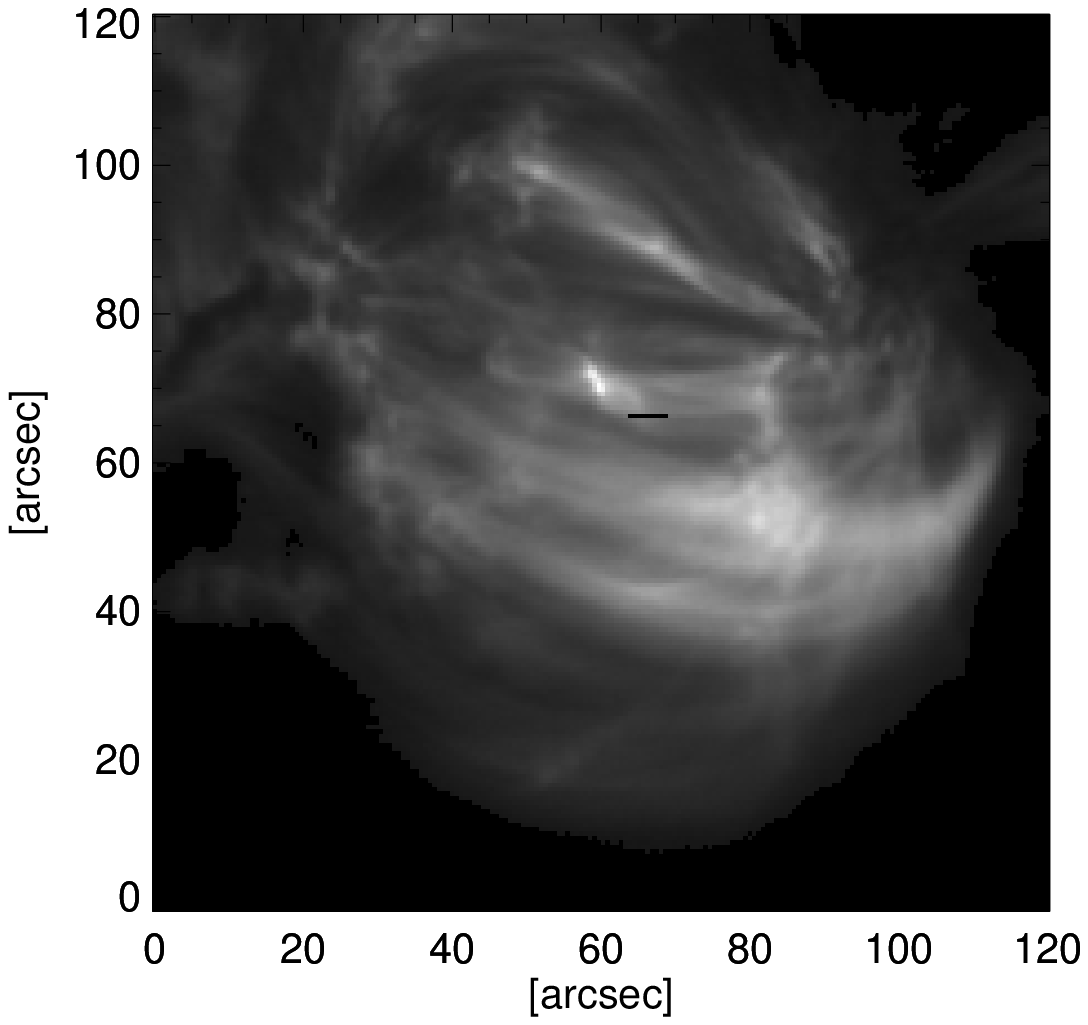}}
\subfigure[]{\includegraphics[width=8cm]{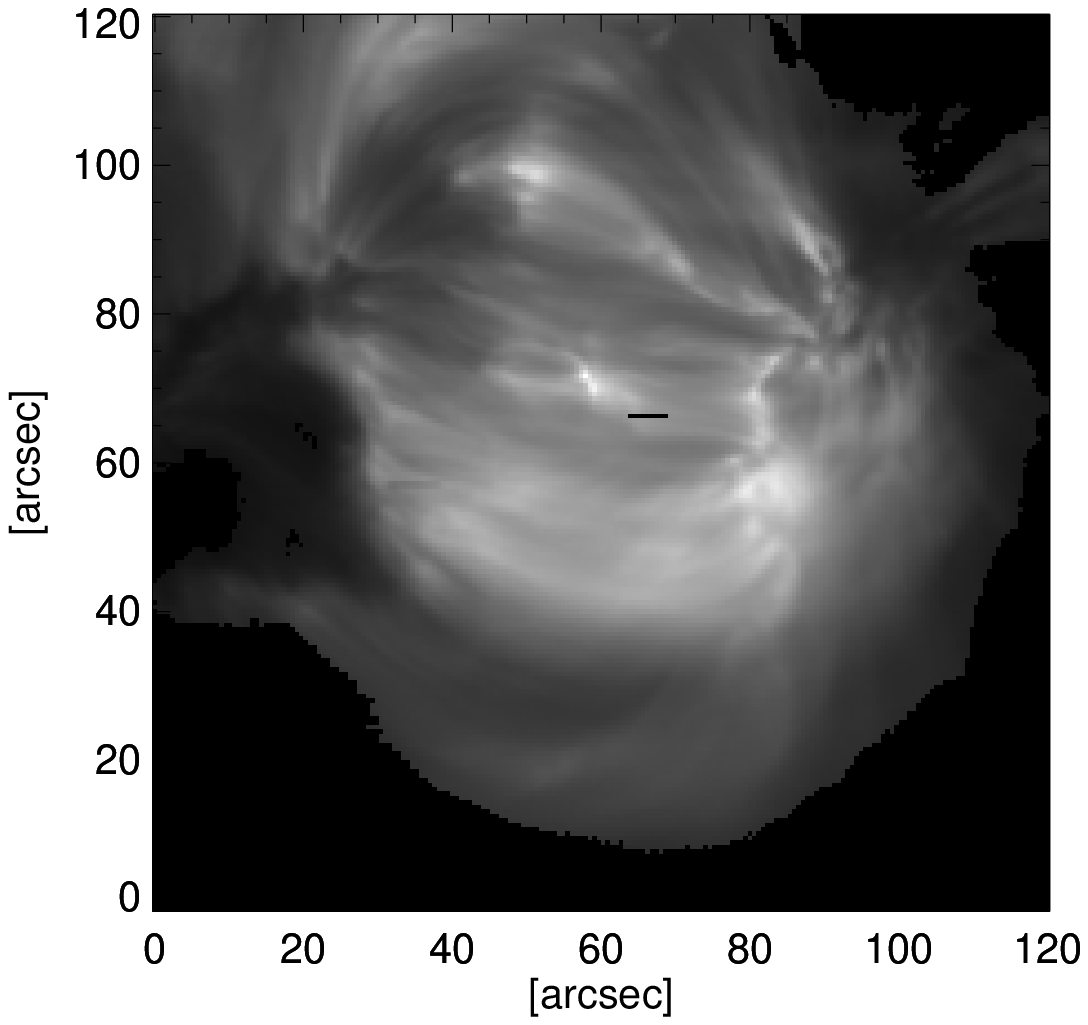}}
\subfigure[]{\includegraphics[width=8cm]{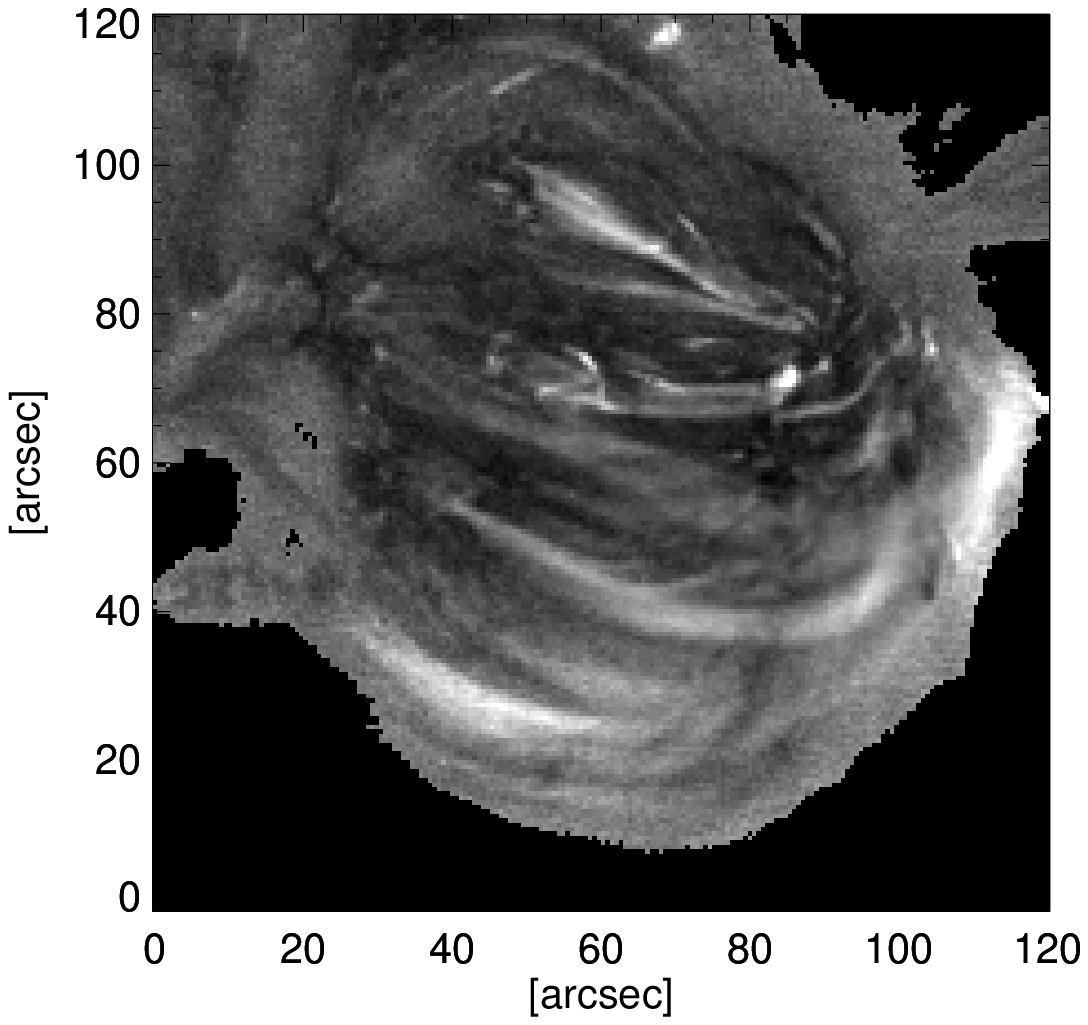}}
\subfigure[]{\includegraphics[width=8cm]{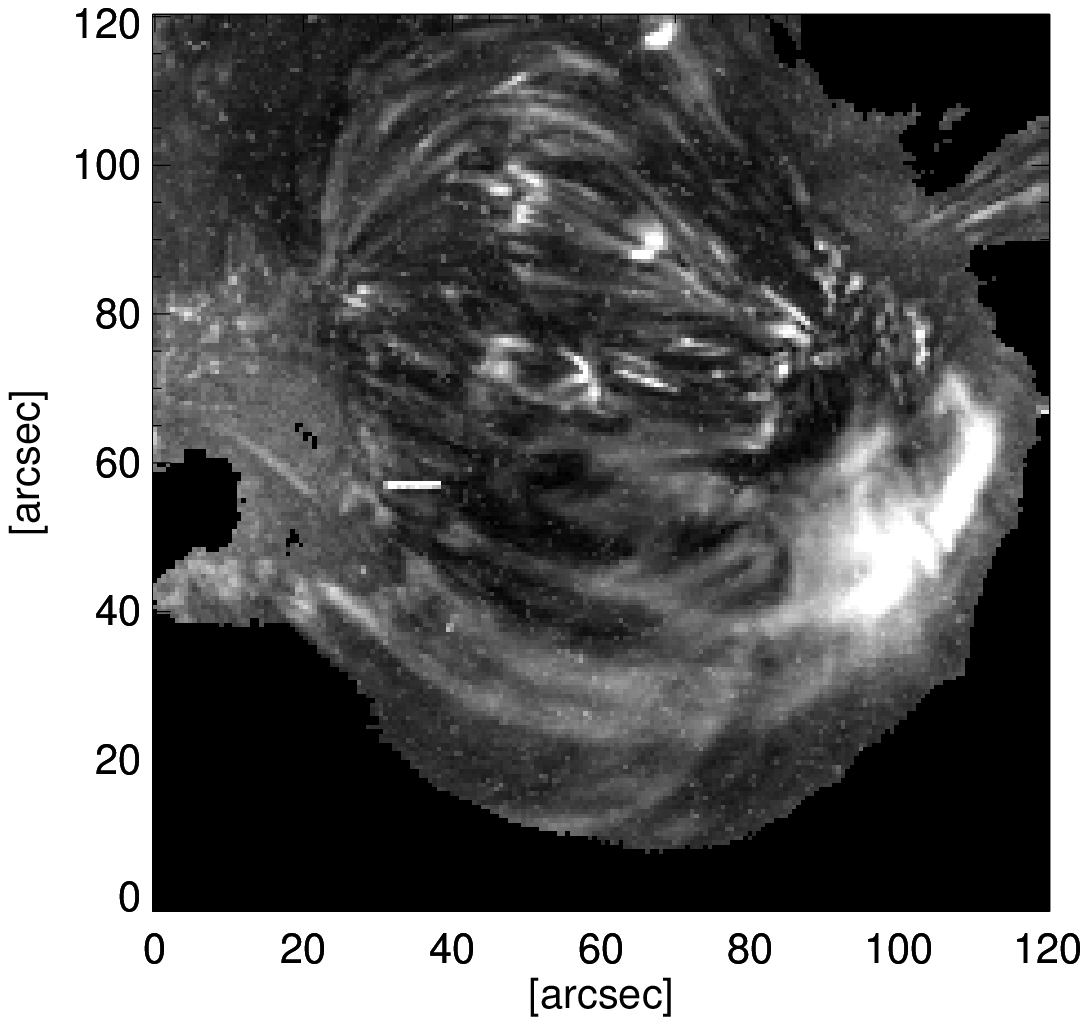}}
% \subfigure[]{\includegraphics[width=8cm]{f94.eps}}
% \subfigure[]{\includegraphics[width=8cm]{f94.eps}}
% \subfigure[]{\includegraphics[width=8cm]{i335.eps}}
 
 \caption{Core of the active region as observed in the 94 \AA\ (a) and 335 \AA\ (b) channels. Only the region above a signal threshold (10 DN s$^{-1}$ pixel$^{-1}$) in the 94 \AA\ channel is shown. The black segments mark the row of pixels whose light curves are shown in Fig.~\ref{fig2}. Maps of standard deviation (normalised to the average) in the 94 \AA\ (c) and 335 \AA\ (d) channel.\aftr{ The grey scales are linear in the range [10\%, 50\%] and [3\%, 20\%], respectively}. In the 335 \AA\ channel the bright row around [X,Y]=[40,60] is not a real feature, but it is a one-pixel spike that drifts rightwards after co-alignment. }
 
\label{fig6}
\end{figure}

We have selected a row of neighbouring pixels among those showing evidence for super-hot plasma in the core of the  active region \citep{Rea11}.  Actually, we have identified a few rows and columns of pixels where the emission shows a coherent time behaviour, i.e. they probably intercept the same loop strands where the confined plasma evolves coherently. As a representative case we consider the same row of pixels as in Paper I. The row contains 9 pixels, i.e. the projected length is $\sim 3.8 \times 10^8$ cm, about 1/4 of the our loop half length projected on the disk. It is not very important to localise exactly which section of the observed loop we are considering, we only ascertain that the loop section taken by the row is all on one side of the loop (left side) with respect to the apex, and does not contain the loop apex. Figure~\ref{fig2} shows the light curves in the 94 \AA\ and 335 \AA\ channels (the third one from the bottom is the single pixel one shown in Paper I). 
As in the model light curves, in each channel the observed light curves are essentially synchronous within our time resolution, with very similar features and bumps that occur at the same time from one pixel to the other.
However, already from a visual inspection, e.g. by comparing the top and bottom light curves, we realise that the amplitude of the fluctuations follow a well-defined trend: they increase going from the bottom to the top in the 94 \AA\ channel, and viceversa in the 335 \AA\ channel. These trends are confirmed quantitatively by comparing the standard deviations of all light curves, which are shown on the right of each curve in Fig.~\ref{fig2}. These trends are the same as those found from our modeling. We have also examined the trends in other randomly selected rows of pixels from the active region core: we do not find opposite trends in the two channels. One might then wonder how general these results are. Fig.~\ref{fig6} is devoted to answer this question. Together with the images of the core of the active region in the 94~\AA\ and 335~\AA\ channel, it shows respective maps of standard deviation $\sigma$ (normalised to the average map for each channel). The $\sigma$-maps are nicely complementary: the body  of many loops are better visible in the 94 \AA\ channel, the footpoints in the 335 \AA\ channel. This confirms larger fluctuations at the footpoints in the ``cooler'' channel and at the apex in the ``hotter'' channel in a significant part of the active region core. The similar emission distribution in the intensity maps of the two channels (Fig.~\ref{fig6}a,b) shows that this effect is not caused by data statistics.

%It is immediate to see that the loop footpoints are mostly green, and the regions in between, i.e. the body of the loops, are mostly red, thus confirming larger fluctuations at the footpoints in the ``cooler'' channel and at the apex in the ``hotter'' channel in a significant part of the active region core.

\section{Conclusions}

Our work shows that a model of finely-structured multi-stranded loop with randomly distributed nanoflares according to a power-law energy distribution predicts larger flickering of pixels near the top of the loop in the 94~\AA\ channel, and near the footpoints in the 335~\AA\ channel, at least in active region cores. The difference is due to the sensitivity of the two channels to different temperature of the emitting plasma, to higher temperature in the former channel (at least in the core of active regions). Therefore, we expect that this effect is not strictly linked to the specific channel, but any bandpasses sensitive to plasma at temperatures roughly above and below 4 MK should show the same trends. The prediction is largely confirmed in the observation of an active region core with evidence for very hot plasma found in \cite{Rea11}, which is a possible signature of impulsive heating.  The presence of fluctuations is intrinsic to the assumption of an impulsive heating finely structured across the loop, which leads to a multi-thermal and variable emission.

The correspondence between large fluctuations at the loop footpoints in the cooler channel and at the loop top in the hotter channel makes the scenario obtained in Paper I further consistent with the observation and also confirms that the heating parameters and the plasma temperatures involved in the model are realistic. 

Therefore, our results support more the model of multi-stranded pulse-heated coronal loop in the active region cores. In the end, we have found easy diagnostics of small scale impulsive heating, which might be useful for extensive application to time sequences of full-disk observations. Further work is necessary to investigate the origin and location of the impulsive heating wherever present.

%%%%%%%%%%\acknowledgments

\acknowledgements{We thank the referee for constructive comments. E.T., F.R.,  and
G.P. acknowledge support from  italian Ministero
dell'Universit\`a e Ricerca. P.T. was supported by contract SP02H1701R from
Lockheed-Martin to the Smithsonian Astrophysical Observatory, and by NASA grant
NNX15AF50G. SDO data supplied courtesy of the SDO/AIA consortia. SDO is the first
mission to be launched for NASA's Living With a Star (LWS) Program.}

\bibliographystyle{aa}
%\bibliography{biblio}

%\begin{thebibliography}{23}
%\expandafter\ifx\csname natexlab\endcsname\relax\def\natexlab#1{#1}\fi

%\end{thebibliography}

\end{document}